# A Continuous Vector-Perturbation for Multi-Antenna Multi-User Communication

Wee Seng Chua, Chau Yuen and Francois Chin

Institute for Infocomm Research (I²R), Singapore
{wschua, cyuen, chinfrancois}@i2r.a-star.edu.sg

*Abstract* — **The sum-rate of the broadcast channel in a multi-antenna multi-user communication system can be achieved by using precoding and adding a regular perturbation to the data vector. The perturbation can be removed by the modulus function, thus transparent to the receiver, but the information of the precoding matrix is needed to decode the symbols. This paper proposes a new technique to improve the multi-antenna multi-user system, by adding a continuous perturbation to the data vector without the need of information on the precoding matrix to be known at the receiver. The perturbation vector will be treated as interference at the receiver, thus it will be transparent to the receiver. The derivation of the continuous vector perturbation is provided by maximizing the signal-to-interference plus noise ratio or minimizing the minimum mean square error of the received signal.**

## I. INTRODUCTION

Given a multi-antenna multi-user communication system, where a base-station with $M$ transmit antennas serving a pool of $K$ autonomous users with one antenna each, the base-station can use two common approaches to approach the sum-rate of the broadcast channel.

The first approach is to multiply a precoding matrix, **G** as shown in (1), to the data vector **u** at the base-station before transmitting. This technique is commonly known as *precoding*, and examples of the precoding matrices included *Inversion* and *Regularized Inversion* [1], the one designed by maximizing sum-rate [2] or minimizing the bit error rate [3] and etc. Besides precoding, the base-station can also add perturbation to the data vector **u** called *vector perturbation* denoted as **v** as shown in (1). It has been shown in [4] that by adding a discrete perturbation vector $\mathbf{v} = \pm \tau \mathbf{l}$, where $\tau$ is a constant value and $\mathbf{l}$ is a vector consisting of only integer value, to the data vector, can reduce the energy of the transmitted signal so as to achieve excellent sum-rate.

$$\mathbf{x} = \frac{\mathbf{G}(\mathbf{u} + \mathbf{v})}{\sqrt{\gamma}} \quad (1)$$

In (1), $\gamma$ is a normalization constant to normalize the total transmission power.

One distinct difference between the two approaches is that the precoding matrix is just a function of the channel but the perturbation vector depends on both the precoding matrix (hence the channel) and the data. While the perturbation remained transparent to the receiver (for the case of vector perturbation in [4], the perturbation will be removed by the modulus function [4] or statistical decoding [5], however, the precoding matrix may or may not be known to the receiver, depending on the assignment of the reference pilot signal. If the pilot signal is precoded with the precoding matrix, the receiver will be able to know the equivalent channel, i.e. **HG**, this is crucial for the decoding of some precoding scheme like regularize-inversion with QAM constellation. However, if the reference signal is precoded, the receiver will not be able to know its raw channel coefficient (i.e. the corresponding row of **H**), this will impose some problems. For example, the raw channel coefficient is important for the receiver to alert the base-station for changing the data rate or precoding matrix.

In this paper, we focus on the case that the reference pilot signal is not precoded, hence inverse precoding will be used because the equivalent channel **HG** of inverse precoding will become an identity matrix, which can be known to the receiver before hand. We propose a new perturbation technique that uses a continuous perturbation vector to improve the performance of multi-antenna multi-user communication system. Moreover, when we combine the continuous perturbation with the discrete perturbation, the performance is better than with only the discrete perturbation alone.

## II. SIGNAL MODEL

The model for the downlink of a multi-antenna multi-user system includes a base-station with $M$ transmit antennas and $K$ users, each with one received antenna can be represented by the following equation

$$\mathbf{y} = \mathbf{H}\mathbf{x} + \mathbf{n} \quad (2)$$

where **y** is a $K \times 1$ column vector with elements representing the received signals for each users, **x** is a $M \times 1$ vector denoting the normalized transmitted signals in (1) with a power constraint $\|\mathbf{x}\|^2 = 1$, **n** represents the additive white Gaussian noise (AWGN) with covariance matrix $\sigma^2 \mathbf{I}$, and **H** represents Rayleigh flat fading channel matrix, and each row of **H**, consists of the corresponding channel of different users.

We use a $1 \times M$ matrix $\mathbf{H}_k^T$ to denote the channel matrix between the base station and the $k$-th user.

The received signal at the $k$-th user is

$$y_k = \mathbf{H}_k^T \mathbf{x} + n_i, \quad k = 1, \ldots, K \tag{3}$$

Based on equations (1) and (2), the transmitted signal $\mathbf{x}$ can be rewritten as

$$\mathbf{y} = \mathbf{H} \frac{\mathbf{G}(\mathbf{u}+\mathbf{v})}{\sqrt{\gamma}} + \mathbf{n} \tag{4}$$

Though in (4), we normalized the transmitted signal instantaneously, it is more practical to normalize the transmitted signal with the expected value of γ, we use the instantaneous normalization for simplicity, as it has been shown in [4] that the difference in performance between two normalization schemes are small.

## III. DISCRETE PERTURBATION

Many optimization algorithms have been proposed in [6], [7], [8] and [9] to maximize the throughput for fixed transmission power, and [10] has provide a comparison between several different schemes.

In [4], the discrete vector-perturbation technique, $\mathbf{v} = \pm \tau l$, is proposed, and it can be implemented using a simplified algorithm called *sphere encoder*. *Sphere encoding* uses a simple and efficient way to achieve capacity of multi-antenna multi-user communication system.

The transmitted signal is

$$\mathbf{x} = \frac{\mathbf{G}(\mathbf{u}+\tau l)}{\sqrt{\gamma}} \tag{5}$$

where $\gamma = \|\mathbf{G}(\mathbf{u}+\tau \mathbf{l})\|^2, \tau = 2(|c|_{\max} + \Delta/2)$

τ is the perturbation interval chosen to provide a symmetric decoding region around every signal constellation point. $|c|_{\max}$ is the absolute value of the largest magnitude among the constellation symbols, and $\Delta$ is the separation between the constellation points.

$l$ is a $K \times 1$ vector whose elements consists of real and imaginary integers. The choice of $l$ is obtained by minimizing γ.

$$l = \arg\min_{l'} (\mathbf{u}+\tau l')^* (\mathbf{H}\mathbf{H}^*)^{-1} (\mathbf{u}+\tau l') \tag{6}$$

Since the value of τ is known to the receiver, and the effect of the integer multiple of τ can be removed by a modulus function.

$$f_\tau(y) = y - \left\lfloor \frac{y+\tau/2}{\tau} \right\rfloor \tau \tag{7}$$

Alternative decoding of the received signal can be found in [5] based on statistic of perturbation.

## IV. CONTINUOUS PERTURBATION

### A. With continuous perturbation only

In this section, we propose a continuous perturbation, where $\mathbf{v} = \mathbf{p}$. It is different from discreet perturbation, where $l$ consists of only integers, $\mathbf{p}$ can be any real or complex value.

When inverse precoding, $\mathbf{G} = \mathbf{H}^*(\mathbf{H}\mathbf{H}^*)^{-1}$, is used, the received signal is

$$\begin{aligned}
\mathbf{y} &= \mathbf{H}\frac{\mathbf{G}(\mathbf{u}+\mathbf{p})}{\sqrt{\gamma}} + \mathbf{n} = \mathbf{H}\frac{\mathbf{H}^*(\mathbf{H}\mathbf{H}^*)^{-1}(\mathbf{u}+\mathbf{p})}{\sqrt{\gamma}} + \mathbf{n} \\
&= \frac{(\mathbf{u}+\mathbf{p})}{\sqrt{\gamma}} + \mathbf{n} \\
&= \underbrace{\frac{\mathbf{u}}{\sqrt{\gamma}}}_{\text{intended signal}} + \underbrace{\frac{\mathbf{p}}{\sqrt{\gamma}}}_{\text{interference}} + \underbrace{\mathbf{n}}_{\text{noise}}
\end{aligned} \tag{8}$$

The first term on the right hand side is the desired signal, the second term is considered interference to the decoder, and the third term is the addictive white Gaussian noise.

The main objective of adding a continuous vector $\mathbf{p}$ to the data vector is to achieve a higher throughput. Thus, an obvious choice of $\mathbf{p}$ is to maximize the SINR

$$SINR = \frac{\|\mathbf{u}\|^2}{\|\mathbf{p}\|^2 + \gamma K \sigma^2} \tag{9}$$

where $K$ refers to the numbers of users.

The normalization constant $\gamma$ is

$$\begin{aligned}
\gamma &= \|\mathbf{G}(\mathbf{u}+\mathbf{p})\|^2 \\
&= \mathbf{u}^*\mathbf{G}^*\mathbf{G}\mathbf{u} + 2\operatorname{Re}(\mathbf{u}^*\mathbf{G}^*\mathbf{G}\mathbf{p}) + \mathbf{p}^*\mathbf{G}^*\mathbf{G}\mathbf{p}
\end{aligned} \tag{10}$$

When (9) and (10) combined, it becomes

$$\begin{aligned}
SINR &= \frac{\|\mathbf{u}\|^2}{\|\mathbf{p}\|^2 + \gamma K \sigma^2} \\
&= \frac{\mathbf{u}^*\mathbf{u}}{\mathbf{p}^*\mathbf{p} + K\sigma^2 \left(\mathbf{u}^*\mathbf{G}^*\mathbf{G}\mathbf{u} + 2\operatorname{Re}(\mathbf{u}^*\mathbf{G}^*\mathbf{G}\mathbf{p}) + \mathbf{p}^*\mathbf{G}^*\mathbf{G}\mathbf{p}\right)} \\
&= \frac{\mathbf{u}^*\mathbf{u}}{D}
\end{aligned} \tag{11}$$

where
$$D = \mathbf{p}^*(\mathbf{I}+K\sigma^2\mathbf{G}^*\mathbf{G})\mathbf{p}+2K\sigma^2\operatorname{Re}(\mathbf{u}^*\mathbf{G}^*\mathbf{Gp})+K\sigma^2\mathbf{u}^*\mathbf{G}^*\mathbf{Gu}$$

Since **p** is continuous, unlike the sphere encoding scheme in [4] which is a discrete value, it can be optimized analytically. Next, we are going to derive **p** by maximizing the SINR, in other words, we take the derivative of (11), we can maximize SINR by minimizing its denominator $D$ by taking the derivative of $D$ with respect to **p**.

$$\frac{\partial D}{\partial \mathbf{p}} = 2(\mathbf{I}+K\sigma^2\mathbf{G}^*\mathbf{G})\mathbf{p}+2K\sigma^2\mathbf{G}^*\mathbf{Gu} \tag{12}$$

Next, we let $\frac{\partial D}{\partial \mathbf{p}} = 0$, to find the optimal **p**

$$\mathbf{p} = -K\sigma^2(\mathbf{I}+K\sigma^2\mathbf{G}^*\mathbf{G})^{-1}\mathbf{G}^*\mathbf{Gu} \tag{13}$$

Similarly, we can minimize the total mean square error of the received signal to find **p**. From (8), the estimate signal is

$$\hat{\mathbf{u}} = \mathbf{u}+\mathbf{p}+\sqrt{\gamma}\mathbf{n} \tag{14}$$

The total mean square error of the received signal is

$$MSE = \|\hat{\mathbf{u}}-\mathbf{u}\|^2 = \|\mathbf{p}+\sqrt{\gamma}\mathbf{n}\|^2 \tag{15}$$

Thus we can find **p** by minimizing (15) by taking its derivative.

$$\frac{\partial MSE}{\partial \mathbf{p}} = 2(\mathbf{I}+K\sigma^2\mathbf{G}^*\mathbf{G})\mathbf{p}+2K\sigma^2\mathbf{G}^*\mathbf{Gu} = 0$$
$$\mathbf{p} = -K\sigma^2(\mathbf{I}+K\sigma^2\mathbf{G}^*\mathbf{G})^{-1}\mathbf{G}^*\mathbf{Gu} \tag{16}$$

It can be seen that (16) is the same as (13), this implies that **p** can be found by using either maximizing the SINR or minimizing the total mean square error of the received signal.

### B. Combine Continuous Perturbation with Discrete Perturbation

In this section, we are going to investigate the effect of continuous perturbation when it is combined with the discrete perturbation. When both the discrete perturbation $\pm \tau l$ and continuous perturbation **p** to the data vector **u**, i.e. $\mathbf{v} = \pm \tau l + \mathbf{p}$, the received signal becomes

$$\mathbf{y} = \mathbf{H}\frac{\mathbf{G}(\mathbf{u}+\tau l+\mathbf{p})}{\sqrt{\gamma}}+\mathbf{n}$$
$$= \frac{(\mathbf{u}+\tau l+\mathbf{p})}{\sqrt{\gamma}}+\mathbf{n} \tag{17}$$
$$= \underbrace{\frac{\mathbf{u}}{\sqrt{\gamma}}}_{\text{intended signal}} + \underbrace{\frac{\tau l}{\sqrt{\gamma}}}_{\text{to be removed by modulus fuction}} + \underbrace{\frac{\mathbf{p}}{\sqrt{\gamma}}}_{\text{interference}} + \underbrace{\mathbf{n}}_{\text{noise}}$$

where $\gamma = \|\mathbf{G}(\mathbf{u}+\tau l+\mathbf{p})\|^2$

The value of $\tau$ is known to the receiver, hence the second term on the right hand side of (17) can be removed by a modulus function [4].

Since the vector perturbation consists of both the continuous and discrete vectors, the cost function of finding $l$ by minimizing $\gamma$ in [4] is no longer valid. It is found contradicting as the value of **p** becomes $-l$ to satisfy the cost function, thus resulting in canceling the discrete perturbation.

Thus, when we find the continuous perturbation, the discrete perturbation can be added to the data vector **u** to find **p**. As a result, the continuous perturbation **p** becomes

$$\mathbf{p} = -K\sigma^2(\mathbf{I}+K\sigma^2\mathbf{G}^*\mathbf{G})^{-1}\mathbf{G}^*\mathbf{G}(\mathbf{u}+\tau l) \tag{18}$$

In this paper, the choice of integer vector $l$ is found concurrently with the continuous vector **p**, which is made with the modified cost function that minimizes the total mean square error of the expected received signal.

From (17), the estimate signal is

$$\hat{\mathbf{u}} = \mathbf{u}+\mathbf{p}+\sqrt{\gamma}\mathbf{n} \tag{19}$$

Hence, the total mean square error of the received signal is

$$MSE = \|\hat{\mathbf{u}}-\mathbf{u}\|^2 = \|\mathbf{p}+\sqrt{\gamma}\mathbf{n}\|^2 \tag{20}$$

The choice of l and p is found by minimizing (20)

$$(\mathbf{p},l) = \arg\min_{\mathbf{p}',l'}\|\mathbf{p}'+\sqrt{\gamma}\mathbf{n}\|^2 \tag{21}$$

Since **p** is given in (18), when we combine (18) and (21), the choice of $l$ becomes

$$l = \arg\min_{l'} \left\| -K\sigma^2(\mathbf{I}+K\sigma^2\mathbf{G}^*\mathbf{G})^{-1}\mathbf{G}^*\mathbf{G}(\mathbf{u}+\tau l') \right.$$
$$\left. + \sqrt{\mathbf{u}^*\mathbf{G}^*\mathbf{Gu}+l'^*\mathbf{G}^*\mathbf{G}l'+\mathbf{p}^*\mathbf{G}^*\mathbf{Gp} \atop +2\operatorname{Re}(\mathbf{u}^*\mathbf{G}^*\mathbf{Gp})+2\operatorname{Re}(\mathbf{u}^*\mathbf{G}^*\mathbf{G}l')+2\operatorname{Re}(l'^*\mathbf{G}^*\mathbf{Gp})}\times\sqrt{K\sigma^2} \right\|^2 \tag{22}$$

Or

$$l = \arg\min_{l'}\|\mathbf{p}+\sqrt{\gamma}\mathbf{n}\|^2$$

where

$$\mathbf{p} = -K\sigma^2(\mathbf{I}+K\sigma^2\mathbf{G}^*\mathbf{G})^{-1}\mathbf{G}^*\mathbf{G}(\mathbf{u}+\tau l) \text{ and}$$
$$\gamma = \mathbf{u}^*\mathbf{G}^*\mathbf{Gu}+l'^*\mathbf{G}^*\mathbf{G}l'+\mathbf{p}^*\mathbf{G}^*\mathbf{Gp}$$
$$+2\operatorname{Re}(\mathbf{u}^*\mathbf{G}^*\mathbf{Gp})+2\operatorname{Re}(\mathbf{u}^*\mathbf{G}^*\mathbf{G}l')+2\operatorname{Re}(l'^*\mathbf{G}^*\mathbf{Gp})$$

## V. SIMULATION RESULTS

Figure 1 and Figure 2 compare inverse precoding with continuous perturbation to the inversion and regularize-inversion precoding without perturbation using uncoded QPSK and 16QAM with $M = K = 4$ respectively.

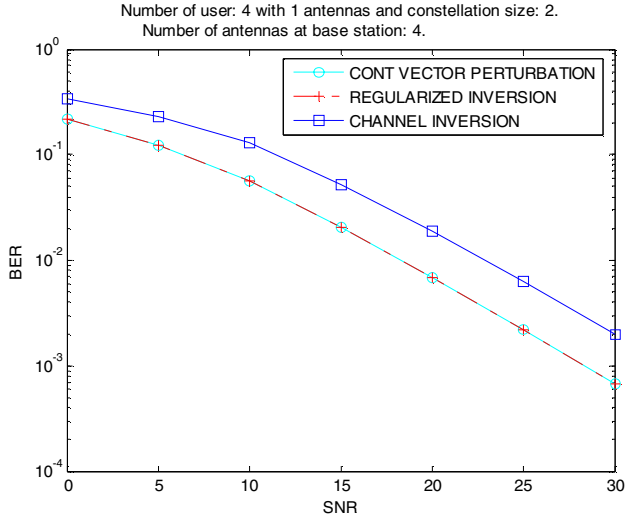

**Figure 1:** Probability of bit error of inverse precoding with continuous perturbation, inverse and regularize-inversion without perturbation using uncoded QPSK symbols, $M=K=4$

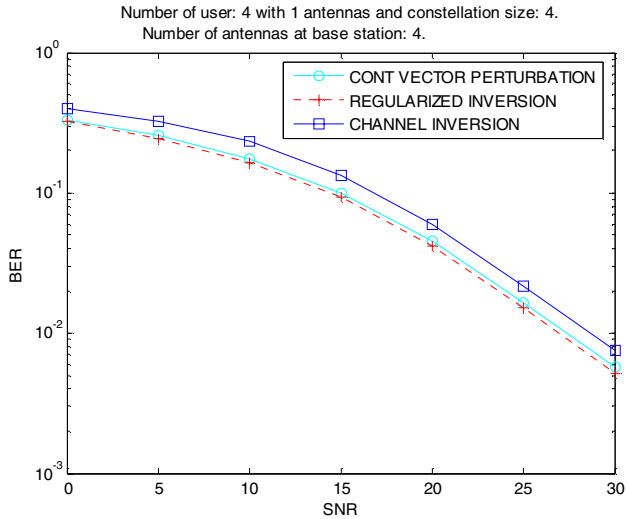

**Figure 2:** Probability of bit error of inverse precoding with continuous perturbation, inverse and regularize-inversion without perturbation using uncoded 16QAM symbols, $M=K=4$

In the case of QPSK, the probability of bit error of the inverse precoding with continuous vector perturbation is identical to regularize-inverse precoding without perturbation. Moreover, it has a 5dB gain over the same precoding without perturbation. Likewise, the difference in probability of bit error between the inverse precoding continuous vector perturbation and the regularize-inverse precoding without perturbation is negligible for 16QAM. However the inversion precoding with continuous perturbation is at least 2dB better than the same precoding without continuous perturbation. It is worth to note that from the results of Figure 1 and Figure 2, continuous perturbation can be used for any constellation symbols.

The comparison of the three techniques using turbo coded 16QAM with $M = K = 4$, using with symbol rate ½ and ¼ respectively is shown in Figure 3 and Figure 4. The difference between the inverse precoding with continuous vector perturbation and the regularize-inverse precoding without any perturbation at turbo coded rate ½ and ¼ is 0.5dB and 1dB respectively. However, inverse precoding with continuous vector perturbation is 1.5dB and 2.5dB better than the same precoding without perturbation at turbo coded rate ½ and ¼ respectively.

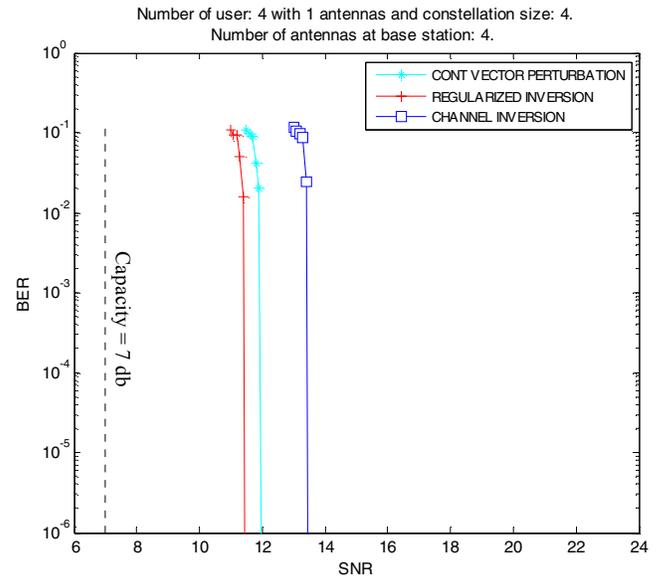

**Figure 3:** Probability of bit error of inverse precoding with continuous perturbation, inverse and regularize-inversion without perturbation using rate ½ turbo coded 16QAM symbols, $M=K=4$

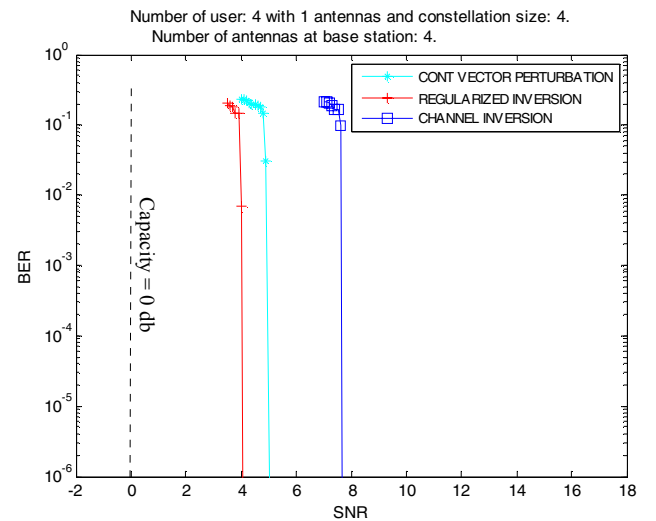

**Figure 4:** Probability of bit error of inverse precoding with continuous perturbation, inverse and regularize-inversion without perturbation using rate ¼ turbo coded 16QAM symbols, $M=K=4$

The results in Figure 5 shows the probability of bit error of inverse precoding with continuous plus discrete perturbation is better than inverse precoding with discrete perturbation and regularize-inverse precoding with discrete perturbation by 1.5dB and 0.5dB respectively.

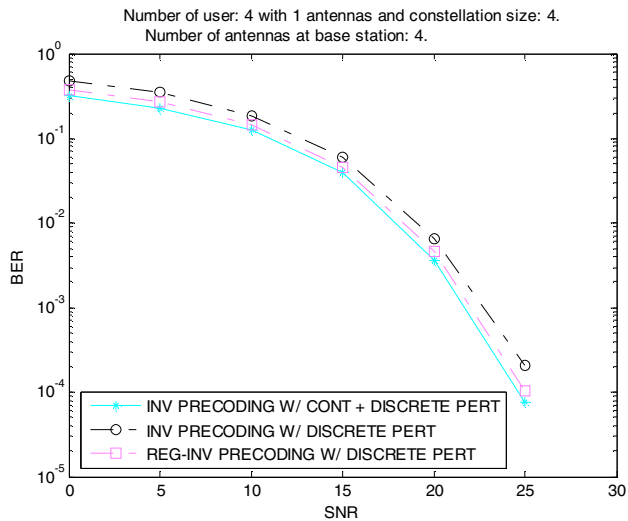

**Figure 5: Probability of bit error of inverse precoding with continuous plus discrete perturbation, inverse and regularize-inversion with discrete perturbation using uncoded 16QAM symbols,** $M=K=4$

## V. CONCLUSIONS

In this paper, we show that by adding a continuous vector perturbation to the data vector, which is treated as interference by the receiver, can achieve a better decoding performance than the system without perturbation. Moreover, the receiver does not need additional training for the precoding matrix.

It is also worth to note that when the continuous perturbation is combined with discrete perturbation, the performance of the inverse precoding continuous plus discrete perturbation is better than inverse precoding or regularize inverse precoding with discrete perturbation only.

## VI. REFERENCES

[1] C. B. Peel, B. M. Hochwald, and A. L. Swindlehurst, "A Vector-Perturbation Technique for Near-Capacity Multi-Antenna Multi-User Communication – Part I: Channel Inversion and Regularization," *IEEE Trans. on Communication.*, vol 53, no. 1, pp. 195-202, Jan 2005.

[2] M. Stojnic, H. Vikalo, and B. Hassibi, "Rate Maximization in Multi-Antenna Broadcast Channels with Linear Preprocessing," *IEEE Trans. on Wireless Communication,* vol 5, issue 9, pp. 2338-2342, Sep 2006.

[3] A. Callard, A. Khandani, and A. Saleh, "Vector Precoding with MMSE for the Fast Fading and Quasi-Static Multi-User Broadcast Channel," 40th Annual Conference on Information Sciences and Systems, pp. 1002-1007, Mar 2006.

[4] B. M. Hochwald, C. B. Peel, and A. L. Swindlehurst, "A Vector-Perturbation Technique for Near-Capacity Multi-Antenna Multi-User Communication – Part I: Perturbation," *IEEE Trans. on Communication.*, vol 53, no. 3, pp. 537-544, Mar 2005.

[5] C. Yuen and B. M. Hochwald, "How to gain 1.5 dB in vector precoding," *IEEE Globecom 2006,* available at http://www.i2r.a-star.edu.sg/~cyuen/publications.html.

[6] Q. H. Spencer, A. L. Swindlehurst, and M. Haardt, "Zero-Forcing methods for downlink spatial multiplexing in multiuser MIMO channels," *IEEE Trans. on Signal Processing*, vol 52, pp. 461-471, Feb. 2004.

[7] M. Schubert and H. Boche, "Iterative multiuser uplink and downlink beamforming under SINR constraints," IEEE Trans. on Signal Processing, vol.53, pp. 2324-2334, Jul 2005.

[8] K. K. Wong, R. D. Murch, and K. B. Letaief, "A joint-channel diagonalization for multiuser MIMO antenna systems," *IEEE Trans. on Wireless Communications*, vol 2, pp. 773-786, July 2003.

[9] D. Samardzija and N. Mandayam, "Multiple antenna transmitter optimization schemes for multiuser systems," *IEEE VTC-Fall 2003*, pp. 399-403.

[10] L. U. Choi, M. T. Ivrlac, R. D. Murch, and W. Utschick, "On strategies of multi-user MIMO transmit signal processing," *IEEE Trans. on Wireless Communications*, vol 3, pp. 1936-1941, Nov 2004.